# Utilizing Low-Cost Sensors to Monitor Indoor Air Quality in Mongolian Gers

Callum E. Flowerday [1], Philip Lundrigan [2], Christopher Kitras [2], Tu Nguyen [3] and Jaron C. Hansen [1,*]

1   Department of Chemistry and Biochemistry, Brigham Young University, Provo, UT 84602, USA
2   Department of Electrical and Computer Engineering, Brigham Young University, Provo, UT 84602, USA
3   Department of Chemistry and Physics, Southeast Missouri State University, One University Plaza, Cape Girardeau, MO 63701, USA; minhtunhatnguyen@gmail.com
*   Correspondence: jhansen@chem.byu.edu; Tel.: +1-801-422-4066

**Abstract:** Air quality has important climate and health effects. There is a need, therefore, to monitor air quality both indoors and outdoors. Methods of measuring air quality should be cost-effective if they are to be used widely, and one such method is low-cost sensors (LCS). This study reports on the use of LCSs in Ulaanbataar, Mongolia to measure $PM_{2.5}$ concentrations inside yurts or "gers". Some of these gers were part of a non-government agency (NGO) initiative to improve insulating properties of these housing structures. The goal of the NGO was to decrease particulate emissions inside the gers; a secondary result was to lower the use of coal and other biomass material. LCSs were installed in gers heated primarily by coal, and interior air quality was measured. Gers that were modified by increasing their insulating capacities showed a 17.5% reduction in $PM_{2.5}$ concentrations, but this is still higher than recommended by health organizations. Gers that were insulated and used a combination of both coal and electricity showed a 19.1% reduction in $PM_{2.5}$ concentrations. Insulated gers that used electricity for both heating and cooking showed a 48% reduction in $PM_{2.5}$ but still had higher concentrations of $PM_{2.5}$ that were 6.4 times higher than recommended by the World Health Organization (WHO). Nighttime and daytime trends followed similar patterns and trends in $PM_{2.5}$ concentrations with slight variations. It was found that at nighttime the outside $PM_{2.5}$ concentrations were generally higher than the inside concentrations of the gers in this study, meaning that $PM_{2.5}$ would flow into the ger whenever the doors were opened, causing spikes in $PM_{2.5}$ concentrations.

## 1. Introduction

The decline of the Earth's air quality has reached a critical level globally, necessitating the implementation of monitoring and mitigation strategies. Air pollutants have a detrimental impact on the health of humans and are currently the fourth leading cause of premature death globally [1]. Particulate matter (PM), specifically $PM_{2.5}$ and $PM_{10}$ with diameters of 2.5 µm and 10 µm, respectively, are commonly monitored sizes of PM. Exposure to PM has been associated with various health conditions, including asthma, cancer, heart disease, type-II diabetes, as well as respiratory and neurodevelopmental disorders [1–3]. In Ulaanbaatar, Mongolia, exposure to $PM_{2.5}$ alone has been identified as the cause of 1400 deaths annually [4]. Lower-income and developing countries, such as Mongolia, are beginning to recognize the adverse impact of air pollution on their population and many are now adopting air quality monitoring practices similar to those used in higher-income countries.

According to the Mongolian Statistical Information Service, as of 2020, approximately 60% of the population in Ulaanbaatar, the capital of Mongolia, reside in areas known as "ger areas". These areas are characterized by the use of coal and wood as primary energy sources. A ger, also known as a yurt, is a traditional round Mongolian hut that can be easily moved to accommodate a nomadic lifestyle. It is noteworthy that 95% of households in these ger areas rely on coal and biomass, such as wood, for cooking and heating. On average, each household burns about 5 tons of coal and approximately 3 m$^3$ of wood per year [5,6].

Ulaanbaatar experiences extreme cold temperatures, the coldest month being January with an average low of −22 °C, making it one of the coldest capital cities in the world. This leads to a significant increase in the burning of coal and biomass during winter months. It is important to note that coal combustion has been identified as a major contributor to particle production, which in turn has adverse effects on health [7,8]. Despite the presence of stacks or chimneys on some coal-burning stoves in these gers, the emissions from coal combustion remain close to the residents' dwellings, resulting in elevated levels of pollutants in the immediate vicinity [9].

Efforts have been made to improve the ventilation and energy efficiency of these stoves [10,11]. These measures, however, have not resulted in significant improvements in indoor air quality [12]. Further studies have been conducted to explore alternative approaches for reducing coal combustion, including the implementation of better insulation techniques for the gers [13–15]. In the past, $PM_{2.5}$ levels in Mongolia have been extremely high, exceeding World Health Organization (WHO) standards by 6.8 times based on 24-h averages [16]. At the central government monitoring site at the United States Embassy, over the 8 months of this study, $PM_{2.5}$ measurements averaged 153 µg/m$^3$, while certain ger neighborhoods experienced levels as high as 229 µg/m$^3$ [17].

To address this issue, the Mongolian government implemented a law on 15 May 2019, prohibiting the use of coal for household consumption. Instead, a program promoting the use of coal briquettes as an alternative fuel source was launched. These efforts produced up to a 50% reduction in outdoor PM levels during the winter months of 2020 in ger neighborhoods, as reported by the National Agency for Meteorology and Environmental Monitoring (NAMEM). Continuous monitoring of indoor PM production, however, and increased assessment frequency are essential.

Monitoring indoor air quality requires a larger number of sensors than outdoor air quality monitoring, which is expensive. To overcome this challenge, low-cost sensors (LCSs) have been developed and utilized, as described in previous studies [18–21]. It is crucial to calibrate these sensors properly, particularly at higher PM concentrations, to account for a phenomenon called "coincidence". Coincidence occurs when the detection of particles is affected by one particle shadowing another, resulting in the second particle scattering no light. Given that most LCSs rely on optical detection, it is vital to mitigate the impact of coincidence.

In our study, we calibrated these LCSs against an optical particle sizer (OPS) with a wide dynamic range, specifically the TSI Model 3300. To minimize the effect of coincidence, the OPS was coupled with a 10:1 diluter. These calibrated LCSs were deployed to measure particulate matter and monitor indoor air quality in rural areas of Mongolia as part of a non-governmental agency (NGO), Deseret International Charities, which sponsored a program to engineer an improved insulation method for gers. Their expectation was that improving the efficiency of the gers would result in lower consumption of coal and biomass resulting in decreased PM emissions and consequently improved outdoor and indoor air quality. The improved insulation method entailed wrapping the circumference and roof of the ger with a radiant barrier coupled with an air gap between the existing layers of felt that are typically used as insulation. A network of 50 LCSs was installed in gers that were heated by coal and electric heat, and air quality was measured before and after the gers were insulated using the engineered solution provided by the NGO.

## 2. Method

### 2.1. Description and Manufacturing of LCSs

There are a variety of LCSs commercially available for a relatively low cost, but a suite of custom air quality sensors was created and used in this study [22,23]. This allowed researchers to have access to the raw data measured by the particulate sensing module in the custom-designed LCS, and additionally none of the commercially available LCSs provided cellular connectivity, which was required for usage in Mongolia. This study's LCS contains a PM sensor (Sensirion $SPS_{30}$), a $CO_2$, temperature and relative humidity sensor (Sensirion $SCD_{30}$), a real-time clock used for timestamps (RTC), an SD card module, and a cellular-enabled microprocessor (Particle Boron 3G/2G). The Sensirion $SCD_{30}$ uses a non-dispersive infrared sensor for $CO_2$ measurements. The particulate matter sensor counts raw measurements of $PM_{2.5}$ by light scattering and calculates the mass of the particles using the refractive index and density of the particle.

As there is no humidity trap on the LCS, a mathematical correction is made to correct for particle size. Though data are uploaded in real-time, all data are logged onto the SD card incorporated with the LCS as a back-up in the event that the sensor loses connectivity with the Internet. The Particle Boron microprocessor coordinates the execution of the sensing modules and data logging. Each LCS is also equipped with a cellular radio that allows for the transmission of data to an online database where the data are stored (InfluxDB) and displayed (Grafana) for analysis. The LCS components are mounted on a printed circuit board (PCB) and encased in a 3D printed box. The total cost of this LCS is approximately USD 200.

### 2.2. Deployment of LCSs

In September 2019, fifty air quality sensors were deployed in gers in Ulaanbaatar. These sensors operated until the end of April 2020. Gers are round, with one doorway, and have a base area of about 11 $m^2$. For this study only data collected from indoor sensors were used. The indoor sensors were placed on one of the center support beams of the ger. Because of a combination of hardware failure, lack of compliance from participants, and difficulty in data retrieval, data were recovered from only 28 sensors. To test the efficacy of new, energy-efficient gers, LCSs were deployed in a variety of conditions: 1 LCS in a modified energy-efficient ger with a coal stove as its source of heat, 3 LCSs in different modified energy-efficient gers with only electric heaters as a source of heat, 19 LCSs in modified energy-efficient gers where both coal stoves and electric heaters were used, and 5 LCSs in traditional, unmodified gers with only coal stoves as a heating source. This last set of gers was used as a control group [24]. The locations of these gers can be seen in Figure 1.

### 2.3. Monitoring of the LCSs

There was a two-pronged approach to LCS maintenance: cloud-based and in-person maintenance. The software solutions include the Particle Cloud Console 2.0, which informed about whether devices were connected to the Internet and uploading data, and the firmware of the sensor, which would notify when the SD card of a sensor was unplugged or not collecting data. In cases of not receiving information from an LCS for an extended period, an affiliate in Ulaanbaatar would visit the site to troubleshoot and repair or replace the sensor.

### 2.4. Data Collection

Data were collected every minute from all the sensors. Every five minutes, data were uploaded to the Internet through the cellular network. If a sensor was not connected to the Internet, it would delay sending data until a connection was established. The backlogged data would then be uploaded. Data were stored in an InfluxDB database located on the campus of Brigham Young University in Provo, Utah, USA. Beyond the measurements of the sensors, metadata were recorded and uploaded, including cell signal strength, software version, and uptime of the sensor.

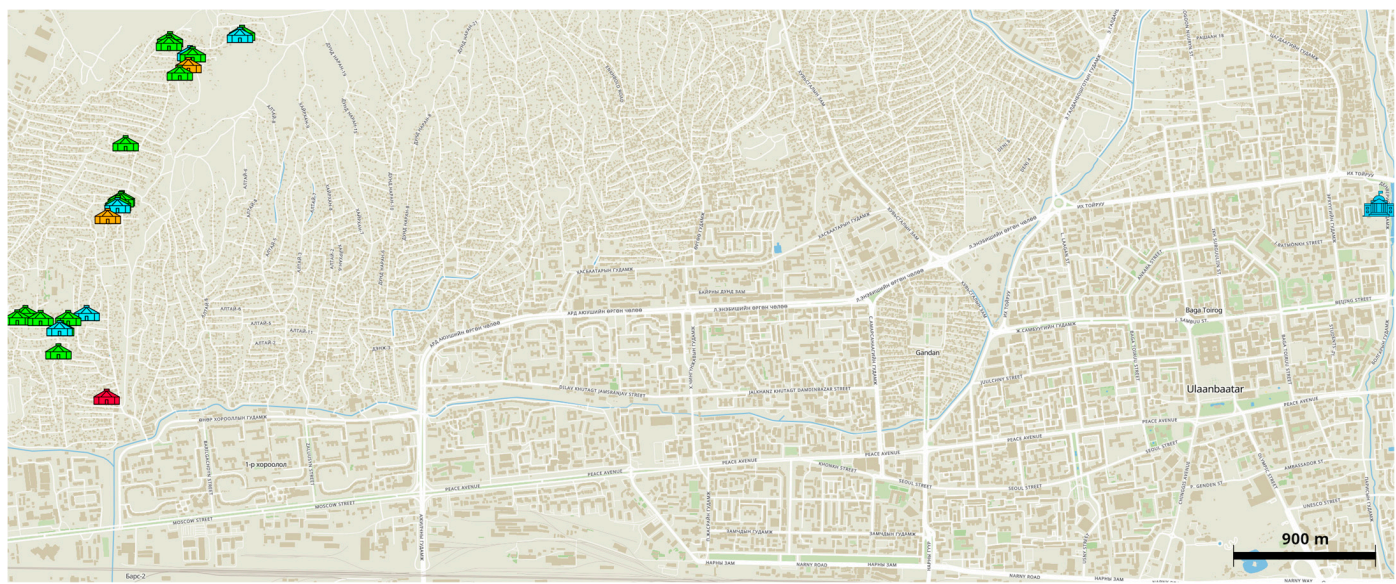

**Figure 1.** Map of Ulaanbaatar, Mongolia showing geographical locations of LCSs in gers and in the U.S. Embassy, which can be seen on the far right in blue. The coal-burning modified ger is in red, the modified electric-only gers in orange, the unmodified gers in blue, and the modified hybrid gers in green.

*2.5. Calibration and Data Correction of LCSs*

2.5.1. Need for Calibration

LCSs are sensitive and can be affected by changes in relative humidity, temperature, sensor aging, and changes in the composition of particulate matter [25–27]. LCSs can also suffer from a deviation of linearity at higher PM concentrations depending on their detection method. For example, if an optical sizing method is being used, some particles may be shadowed by other particles and the total PM count will be biased low in its reported concentration. These conditions can be corrected for by calibrating the LCSs against reference PM monitors, through modelling, or collocating monitors. We have calibrated our Sensirion LCSs against an optical particle sizer (TSI Optical Particle Sizer model 3300) with a 1:10 diluter (TSI Particle Diluter Model 3332). The error in the TSI OPS is 5% at 0.5 µm, inherent with the error in the flowrate through the instrument. When compared with other LCSs, the Sensirion was one of the more stable LCSs over time and in many environments [28]. The Sensirion LCS can work in 0–95% humidity and in the temperature range of $-10\,°C$ to $60\,°C$ without correction. The age limit of the sensor is recommended to be 8 years by the manufacturer; however, aging can happen over time. Corrections for aging were made in second-order polynomial regression as this affects larger concentrations more than smaller concentrations of $PM_{2.5}$. Our LCSs were calibrated after 6 months of use to minimize the effects of aging.

2.5.2. Calibration of the LCSs

The experimental setup for calibration can be seen in Figure 2. Calibration of an LCS was carried out by attaching the LCS to the roof of our atmospheric chamber, made from a 55 gallon drum, and having the TSI OPS draw air from the top of the atmospheric chamber through the diluter. The chamber was cleaned out with the use of T8 UV lights and zero air for several hours before calibration in order to photolyze, dilute, and remove leftover $PM_{2.5}$. The diluter was used to minimize the effects of coincidence with the TSI OPS, which already had a higher tolerance range than the LCS. A cast iron stove (Regency Fireplace

Products), similar in design to what is used in Mongolian gers, was used to simulate the conditions in the gers. Anthracite coal, like that burned in Mongolia, was burned in the stove, and smoke from the stove was directed into the environmental chamber. Calibration of the LCS was carried out using smoke from a source like that present in Mongolia. An extractor fan (Antrader) was used to pull smoke from directly above the chimney of the coal-burning stove through a 1″ diameter PVC tube into the bottom of the reaction chamber. Another 1″ diameter fan was mounted at the exit of the PVC pipe that rested on the bottom of the chamber, and a final 1″ diameter fan was mounted 33″ up the wall of the chamber opposite to the exit of the pipe. These two fans were used to ensure mixing of the gases in the barrel.

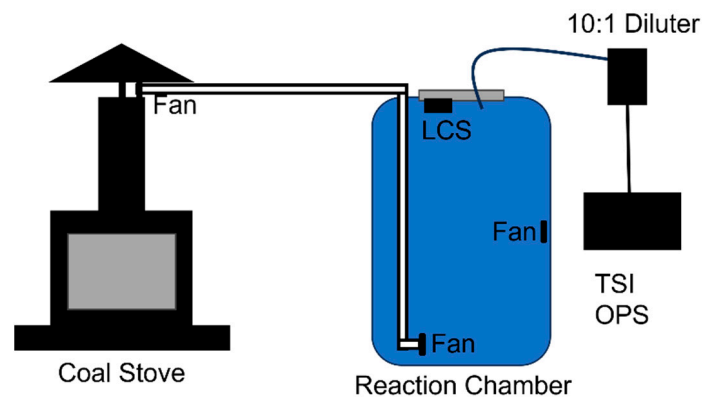

**Figure 2.** Schematic of experimental setup for calibration.

Data from the LCS in the barrel were plotted against the data from the TSI OPS for the same period. This yielded a second-order polynomial transfer function seen in Figure 3. The transfer function was seen to affect larger values of concentrations of $PM_{2.5}$ more than smaller ones. This is due to the increased effect of coincidence in the detection of $PM_{2.5}$ at these higher concentrations. This second-order polynomial regression yielded a line formula that was used to correct the LCS data collected in Mongolia.

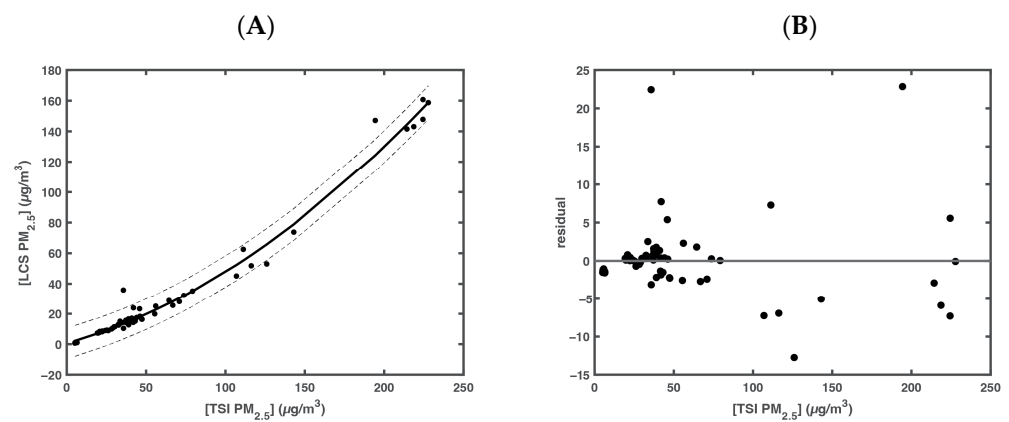

**Figure 3.** (**A**) LCS data plotted against the TSI OPS data from the transfer function run. Trend line to data is plotted as a solid black line. The 95% confidence intervals to the fit is plotted as dashed lines. (**B**) Residual plot of fit to data.

Table 1 shows a summary of the number of data points and sensors from each class of ger in this study. Data from each of the four classes of gers, unmodified, modified coal-burning, modified hybrid of coal and electric, and modified electric only, were baseline corrected, where necessary, and averaged to produce a single dataset for each of the conditions. Standard deviations and confidence intervals were then calculated from these averaged datasets. A Grubb's Test, at 95%, was carried out on all sensors to exclude outliers.

Datasets were adjusted accordingly before continuing. Daytime and nighttime collected data were separated for analysis. Daytime hours were defined as 5 am–8 pm (local time) and were analyzed separately from the nighttime collected data to deconvolve the influence of cooking activities from heating. Daytime hours were defined based on an evaluation of the diurnal pattern of cooking, as observed in Figure 4. Figure 4 shows typical $PM_{2.5}$ concentrations of <5 µg m$^{-3}$ between the hours of 12:00 pm and 5 am in an unmodified ger. Typically, beginning at about 5 am, the $PM_{2.5}$ concentrations begin to rise as meal preparation begins. Three peaks are usually observed between the hours of 5 am and 8 pm associated with meal preparation throughout the day. $PM_{2.5}$ concentrations typically begin to decrease after the final meal preparation in the evening and slowly decay to baseline conditions as the evening progresses.

**Table 1.** Visualization of raw sensor data breakdown.

| Ger Type | Number of Sensors | Number of Averaged Data Points |
| --- | --- | --- |
| Unmodified | 6 | 187,456 |
| Coal-only modified | 1 | 448,233 |
| Hybrid modified | 20 | 110,139 |
| Electric only | 3 | 583,460 |

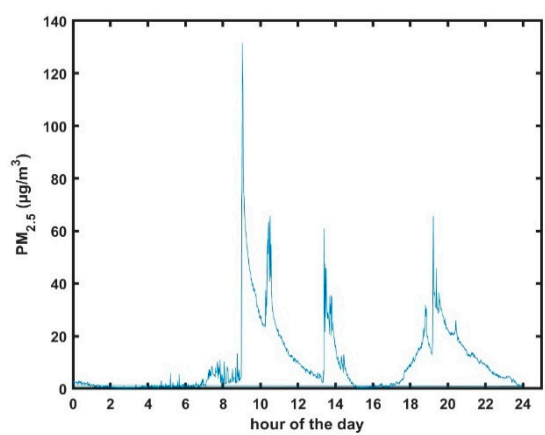

**Figure 4.** Example of the diurnal pattern in $PM_{2.5}$ concentrations showing cooking spikes for breakfast, lunch, and dinner.

## 3. Results and Discussion

### 3.1. Averaged Data

Figure 5 shows the average $PM_{2.5}$ values for each class of ger for the duration of the study. A pattern is observed that shows that, as the use of coal decreases, the $PM_{2.5}$ concentration decreases. The highest concentration of $PM_{2.5}$ is found in the unmodified, $187.0 \pm 1.0$ µg/m$^3$, and modified but coal-burning, $154.3 \pm 0.5$ µg/m$^3$, gers. It is expected that the modified gers will burn less coal as the added insulation traps more heat; however, by increasing the insulating value of the structure, it also traps $PM_{2.5}$ in the ger by impeding the venting of smoke. Despite the possible ventilation inhibition caused by the added insulation, it does appear to be favorable to insulate the gers as the total $PM_{2.5}$ concentrations drop by 32.7 µg/m$^3$ or 17.5%. The third highest $PM_{2.5}$ concentration, $151.4 \pm 0.8$ µg/m$^3$, is found in the modified hybrid ger. $PM_{2.5}$ dropped on average by 35.6 µg/m$^3$ or 19.1% relative to the unmodified ger. The addition of electric heating to the ger lowered the $PM_{2.5}$ produced from the burning of coal. The increased insulation helped to preserve the heat in the structure and reduce the amount of $PM_{2.5}$ being produced by the ger and released into the environment due to decreased heating needs. The lowest $PM_{2.5}$ concentration was found in the modified electric ger, at $95.5 \pm 0.5$ µg/m$^3$. This is expected as there is little to no coal burning in these gers, eliminating this major source of $PM_{2.5}$. $PM_{2.5}$ dropped on average by 91.5 µg/m³ or 48.9% relative to the unmodified ger. It should be noted that in the modified electric ger, the average total $PM_{2.5}$ limits are still 6.4 times higher than the WHO standards of 15 µg/m³ [29]. This is consistent with the findings of other studies [16]. Outside air quality data came from sensors outside the U.S. Embassy in Ulaanbaatar. The outside $PM_{2.5}$ concentrations for the period of this study averaged 153.5 ± 4.6 µg/m³, which is 10.2 times the WHO standards. Outside air quality data were used to show that air inside the gers is cleaner than air outside the gers. This justifies the conclusion that spikes in $PM_{2.5}$ concentrations are seen when entrances to the gers are opened, as $PM_{2.5}$ diffuses from high concentrations outside the gers to low concentrations inside the gers.

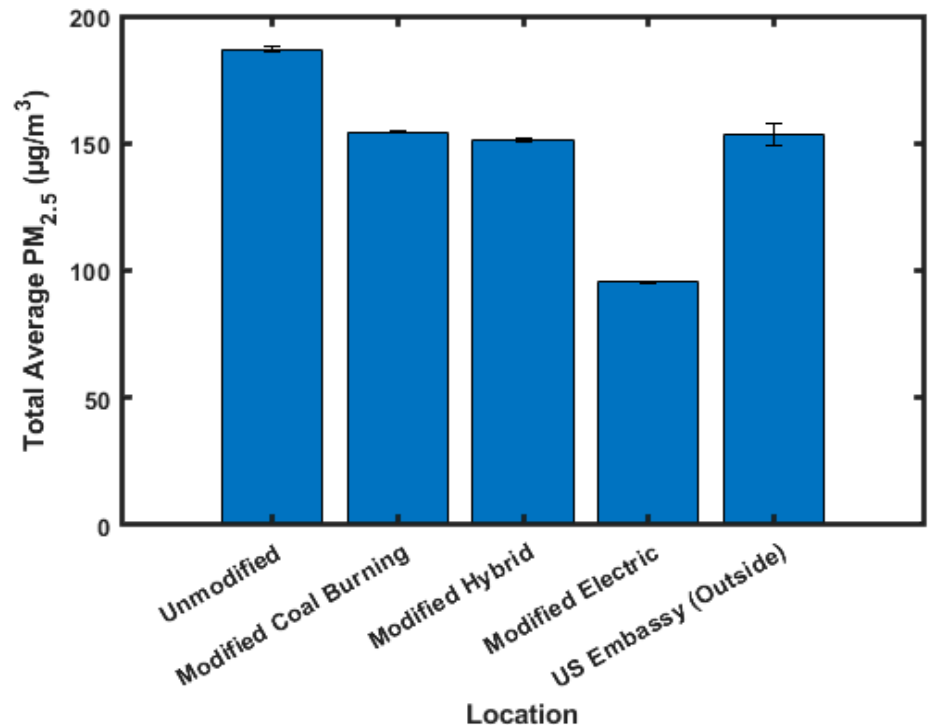

**Figure 5.** Average $PM_{2.5}$ values.

### 3.2. Daily and Nightly Averaged Data

Figure 6 shows the averaged daytime, 5 am–8 pm, $PM_{2.5}$ values, and Figure 7 shows the averaged nighttime, 8 pm–5 am, $PM_{2.5}$ values. These times were chosen to include cooking-related $PM_{2.5}$ spikes included in the daytime data, and so that nighttime data are primarily influenced from heating.

Daytime values show the highest $PM_{2.5}$ concentration in the unmodified (180.3 ± 1.3 µg/m³) ger, followed by the modified hybrid (142.0 ± 1.1 µg/m³), modified coal-burning (126.2 ± 0.5 µg/m³), and modified electric (84.4 ± 0.6 µg/m³) gers. The nighttime data show a slightly different pattern compared to the daytime and averaged data, with the highest $PM_{2.5}$ concentration being in the modified coal-burning (200.5 ± 0.7 µg/m³) ger, followed by the unmodified (198.3 ± 1.3 µg/m³), modified hybrid (167.5 ± 1.0 µg/m³), and modified electric (113.4 ± 1.0 µg/m³) gers. It should be noted that there was only one LCS located in a modified coal-burning ger, and that the high nighttime concentration measured could be caused by either the increased insulation in this ger trapping $PM_{2.5}$ or an artifact of the proximity of other gers that were burning coal that could contribute to its higher $PM_{2.5}$ concentrations. The daytime outside $PM_{2.5}$ concentrations, 108.4 ± 4.8 µg/m³, are lower than all indoor average values besides modified electric gers. The nighttime outside $PM_{2.5}$ concentrations, 229.0 ± 7.9 µg/m³, are higher than all indoor averages except for modified coal-burning gers. This means that when the doors are opened to these gers, a rush of PM$_{2.5}$ is introduced into the gers increasing the indoor PM$_{2.5}$ concentration. This bias may be magnified in the modified gers as the insulation will trap more of the PM$_{2.5}$.

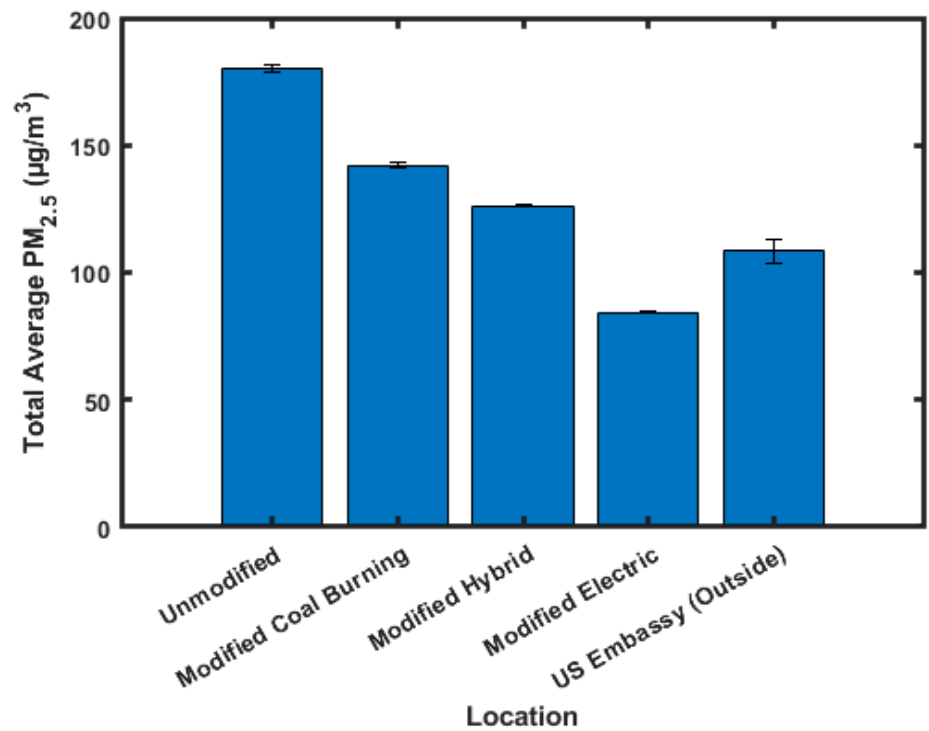

**Figure 6.** Averaged daytime PM$_{2.5}$ values.

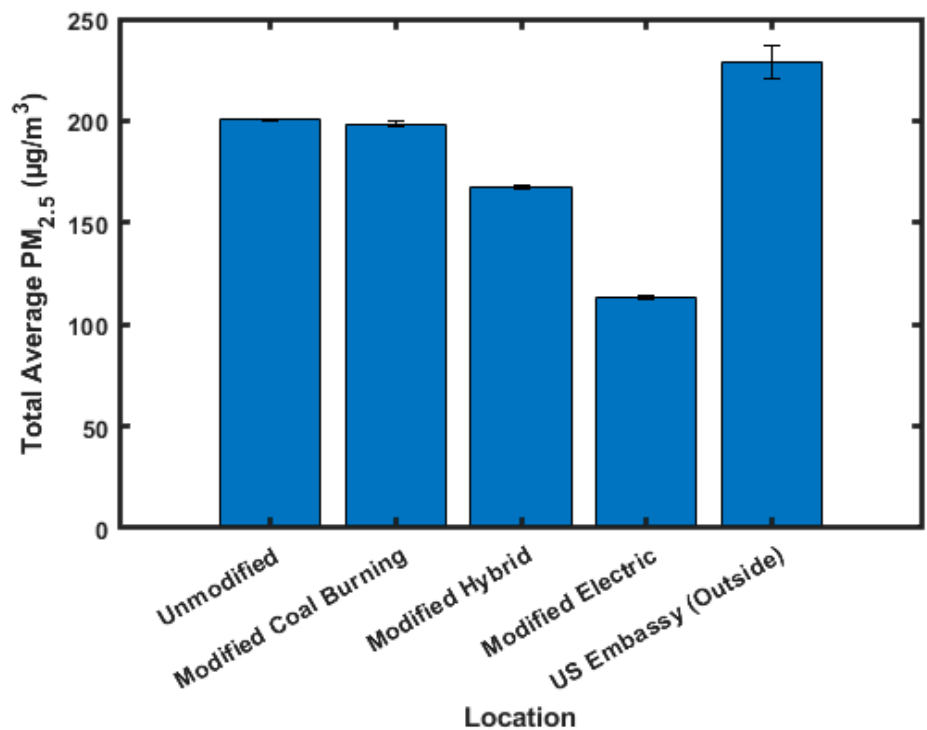

**Figure 7.** Averaged nighttime PM$_{2.5}$ values.

## 4. Conclusions

Efforts are underway to improve air quality in third-world countries like Mongolia, particularly by addressing the issue of pollution caused by coal burning. The approach implemented in this study was to insulate gers in an attempt to reduce coal consumption and subsequently lower indoor air $PM_{2.5}$ concentrations. The hypothesis tested was that by increasing the insulating value of gers, less coal would be burned resulting in lower indoor $PM_{2.5}$ concentrations. This study found that overall $PM_{2.5}$ concentrations were 17.5% lower in gers that were insulated. We do acknowledge that whenever the doors to the gers were opened, it is possible that $PM_{2.5}$ from surrounding gers and the outside environment came into the modified gers and that this may bias the data. This seemed more evident at nighttime when there was more activity in the gers. The reduction in ventilation caused by the insulation appears to trap $PM_{2.5}$ in the ger for longer periods before the $PM_{2.5}$ can vent out of the ger. Despite the effect of increasing insulation reducing ventilation of $PM_{2.5}$, it is recommended that gers be insulated to reduce coal consumption and to decrease $PM_{2.5}$ exposure. We also hypothesize that if neighboring gers were also insulated, the total $PM_{2.5}$ concentrations in the outdoor air may be lowered and that this would also have the effect of increasing indoor air quality. This study shows that the most effective approach for lowering indoor $PM_{2.5}$ concentrations is to use electricity for both cooking and heating in gers. The data show that gers that used electricity had the lowest $PM_{2.5}$ concentrations. $PM_{2.5}$ dropped on average by 48.9%. For example, the hybrid coal and electric approach showed an increase in air quality compared to the coal-burning-only gers, but it was not as effective at lowering $PM_{2.5}$ concentrations as a fully electrified ger. It is important to note that the fully electrified ger still did not obtain the $PM_{2.5}$ standard set by the NAAQS. This may be because gers surrounding the fully electric gers continued to burn coal, and these emissions likely leaked into the electrified-only gers. While it is recognized that access to electricity is not always feasible in Mongolia, it is recommended that electricity be used to lower indoor $PM_{2.5}$ concentrations where possible.

**Author Contributions:** Conceptualization, P.L. and J.C.H.; Formal analysis, C.E.F., C.K., T.N. and J.C.H.; Data curation, P.L., C.K., C.E.F. and J.C.H.; Writing—original draft, C.E.F.; Writing—review & editing, P.L., C.K. and J.C.H.; Project administration, J.C.H.; Funding acquisition, P.L. All authors have read and agreed to the published version of the manuscript.

**Funding:** Deseret International Charities in Salt Lake City #R0702018, Utah, USA and through a generous donation from Milt and Heidi Shipp.

**Institutional Review Board Statement:** Not applicable.

**Informed Consent Statement:** Not applicable.

**Data Availability Statement:** Data are available at https://scholarsarchive.byu.edu/facpub/6973 (accessed on 6 July 2023).

**Conflicts of Interest:** The authors declare no conflict of interest.

## References

1. Institute, H.E. *State of Global Air 2020*; Special Report; Health Effects Institute: Boston, MA, USA, 2020.
2. Jadambaa, A.; Spickett, J.; Badrakh, B.; Norman, R.E. The Impact of the Environment on Health in Mongolia: A Systematic Review. *Asia-Pac. J. Public Health* **2015**, *27*, 45–75. [CrossRef]
3. Arvanitis, A.; Kotzias, D.; Kephalopoulos, S.; Carrer, P.; Cavallo, D.; Cesaroni, G.; De Brouwere, K.; de Oliveira-Fernandes, E.; Forastiere, F.; Fossati, S.; et al. The Index-Pm Project: Health Risks from Exposure to Indoor Particulate Matter. *Fresenius Environ. Bull.* **2010**, *19*, 2458–2471.
4. Hill, L.D.; Edwards, R.; Turner, J.R.; Argo, Y.D.; Olkhanud, P.B.; Odsuren, M.; Guttikunda, S.; Ochir, C.; Smith, K.R. Health assessment of future PM2.5 exposures from indoor, outdoor, and secondhand tobacco smoke concentrations under alternative policy pathways in Ulaanbaatar, Mongolia. *PLoS ONE* **2017**, *12*, e0186834. [CrossRef] [PubMed]
5. World Bank. *Mongolia—Energy Efficient and Cleaner Heating in Poor, Peri-Urban Areas of Ilaanbaatar: Summary Report on Activies*; World Bank: Washington, DC, USA, 2008.
6. Guttikunda, S. *Urban Air Pollution Analysis for Ulaanbaatar*; Ministry of the Environment and Green Development: Ulaanbaatar, Mongolia, 2007.
7. Davy, P.K.; Gunchin, G.; Markwitz, A.; Trompetter, W.J.; Barry, B.J.; Shagjjamba, D.; Lodoysamba, S. Air particulate matter pollution in Ulaanbaatar, Mongolia: Determination of composition, source contributions and source locations. *Atmos. Pollut. Res.* **2011**, *2*, 126–137. [CrossRef]

8. Nakao, M.; Yamauchi, K.; Ishihara, Y.; Omori, H.; Ichinnorov, D.; Solongo, B. Effects of air pollution and seasons on health-related quality of life of Mongolian adults living in Ulaanbaatar: Cross-sectional studies. *BMC Public Health* **2017**, *17*, 594. [CrossRef] [PubMed]
9. World Bank. *Air Pollution in Ulaanbaatar: Initial Assessment of Current Situation and Effects of Aabatement Measures*; World Bank: Washington, DC, USA, 2009.
10. Lodoyasamba, S.; Pemberton-Pigott, C. Mitigation of Ulaanbaatar city's air pollution—From source apportionment to ultra-low emission lignite burning stoves. *New Dawn Eng.* **2011**.
11. Greene, L.; Turner, J.; Edwards, R.; Cutler, N.; Duthie, M.; Rostapshova, O. *Impact Evaluation Results of the MCA Mongolia Energy and Environment Project Energy-Efficient Stove Subsidy Program*; Milleniumn Challenge Corporation: Washington, DC, USA, 2014.
12. Cowlin, S.; Kaufmann, R.B.; Edwards, R.; Smith, K.R. *Impact of Improved Stoves on Indoor Air Quality in Ulaanbaatart, Mongolia*; World Bank: Washington, DC, USA, 2005.
13. Lim, M.; Myagmarchuluun, S.; Ban, H.; Hwang, Y.; Ochir, C.; Lodoisamba, D.; Lee, K. Characteristics of Indoor PM2.5 Concentration in Gers Using Coal Stoves in Ulaanbaatar, Mongolia. *Int. J. Environ. Res. Public Health* **2018**, *15*, 2524. [CrossRef] [PubMed]
14. Ezzati, M.; Kammen, D.M. The health impacts of exposure to indoor air pollution from solid fuels in developing countries: Knowledge, gaps, and data needs. *Environ. Health Perspect.* **2002**, *110*, 1057–1068. [CrossRef] [PubMed]
15. Soyol-Erdene, T.O.; Ganbat, G.; Baldorj, B. Urban Air Quality Studies in Mongolia: Pollution Characteristics and Future Research Needs. *Aerosol Air Qual. Res.* **2021**, *21*, 210163. [CrossRef]
16. Wang, M.R.; Kai, K.; Sugimoto, N.; Enkhmaa, S. Meteorological Factors Affecting Winter Particulate Air Pollution in Ulaanbaatar from 2008 to 2016. *Asian J. Atmos. Environ.* **2018**, *12*, 244–254. [CrossRef]
17. US Embassy Ulaanbaatar Mongolia. Ulaanbaatar—Historical Data. 2023. Available online: https://stateair.mn/history.php (accessed on 6 July 2023).
18. Raheja, G.; Sabi, K.; Sonla, H.; Gbedjangni, E.K.; McFarlane, C.M.; Hodoli, C.G.; Westervelt, D.M. A Network of Field-Calibrated Low-Cost Sensor Measurements of PM$_{2.5}$ in Lomé, Togo, Over One to Two Years. *ACS Earth Space Chem.* **2022**, *6*, 1011–1021. [CrossRef] [PubMed]
19. Hegde, S.; Min, K.T.; Moore, J.; Lundrigan, P.; Patwari, N.; Collingwood, S.; Balch, A.; Kelly, K.E. Indoor Household Particulate Matter Measurements Using a Network of Low-cost Sensors. *Aerosol Air Qual. Res.* **2020**, *20*, 381–394. [CrossRef]
20. Min, K.T.; Lundrigan, P.; Patwari, N. IASA—Indoor air quality sensing and automation: Demo abstract. In Proceedings of the 16th ACM/IEEE International Conference on Information Processing in Sensor Networks, Pittsburgh, PA, USA, 18–21 April 2017; Association for Computing Machinery: Pittsburgh, PA, USA, 2017; pp. 277–278.
21. Moore, J.; Goffin, P.; Meyer, M.; Lundrigan, P.; Patwari, N.; Sward, K.; Wiese, J. Managing In-home Environments through Sensing, Annotating, and Visualizing Air Quality Data. *Proc. ACM Interact. Mob. Wearable Ubiquitous Technol.* **2018**, *2*, 128. [CrossRef]
22. PurpleAir. PurpleAir: Real-Time Air Quality Monitoring. 2022. Available online: https://www2.purpleair.com/ (accessed on 6 July 2023).
23. AirU. AirU at the University of Utah. 2017. Available online: https://airu.coe.utah.edu/about/ (accessed on 6 July 2023).
24. Alcantara, L.; Miera, J.; Ariun-Erdene, B.; Teng, C.C.; Lundrigan, P. The Hitchhiker's Guide to Successful Remote Sensing Deployments in Mongolia. In Proceedings of the 2020 Intermountain Engineering, Technology and Computing (IETC), Orem, UT, USA, 2–3 October 2020; pp. 1–6.
25. RoRivero, A.G.; Rivero, R.A.G.; Schalm, O.; Rodríguez, E.H.; Rodríguez, E.H.; Pérez, M.C.M.; Caraballo, V.N.; Jacobs, W.; Laguardia, A.M. A Low-Cost Calibration Method for Temperature, Relative Humidity, and Carbon Dioxide Sensors Used in Air Quality Monitoring Systems. *Atmosphere* **2023**, *14*, 191. [CrossRef]
26. Nieckarz, Z.; Zoladz, J.A. New Calibration System for Low-Cost Suspended Particulate Matter Sensors with Controlled Air Speed, Temperature and Humidity. *Sensors* **2021**, *21*, 5845. [CrossRef] [PubMed]
27. Lee, H.; Kang, J.; Kim, S.; Im, Y.; Yoo, S.; Lee, D. Long-Term Evaluation and Calibration of Low-Cost Particulate Matter (PM) Sensor. *Sensors* **2020**, *20*, 3617. [CrossRef]
28. Tryner, J.; Mehaffy, J.; Miller-Lionberg, D.; Volckens, J. Effects of aerosol type and simulated aging on performance of low-cost PM sensors. *J. Aerosol Sci.* **2020**, *150*, 105654. [CrossRef]
29. WHO. What Are the WHO Air Quality Guidelines? 2021. Available online: https://www.who.int/news-room/feature-stories/detail/what-are-the-who-air-quality-guidelines (accessed on 6 April 2023).